\documentclass[preprint,showpacs,preprintnumbers,amsmath,amssymb]{revtex4-1}
\usepackage{mathrsfs}
\usepackage{amsmath}
\usepackage{mathtools}
\usepackage{graphicx}
\usepackage{footnote}
\usepackage{verbatim}
\usepackage{epstopdf}
\usepackage[usenames]{color}
\usepackage{graphics}
\usepackage{bm}
\usepackage{amsmath}
\usepackage{dcolumn} 
\usepackage{tikz}
\usepackage{rotating}
\usepackage{graphics}
\usepackage{epsfig}
\begin{document}
\input{epsf}
\preprint{APS/12three-QED}
\title{Effect of strong $\bar{\rm p}$-p nuclear forces on the rate of the
low-energy three-body protonium formation reaction:
$\bar{\mbox{p}} + \mbox{H}_{\mu}(1s) \rightarrow (\bar{\mbox{p}} \mbox{p})_{\alpha} + \mu^-$}
%
\author{Renat A. Sultanov$^{1,2}$\footnote{Electronic mail: rasultanov@stcloudstate.edu;\ r.sultanov2@yahoo.com}}
\author{Dennis Guster$^{2}$\footnote{Electronic mail: dcguster@stcloudstate.edu}}
\author{Sadhan K. Adhikari$^{1}$\footnote{Electronic mail: adhikari@ift.unesp.br;\ http://www.ift.unesp.br/users/adhikari}}
\affiliation{$^{1)}$Instituto de F\'isica Te\'orica, UNESP $-$ Universidade Estadual Paulista, 01140 S\~ao Paulo, SP, Brazil\\
$^{2)}$Department of Information Systems, BCRL \& Integrated Science and Engineering Laboratory
Facility \it{(ISELF)} at St. Cloud State University,  St. Cloud, MN, USA}

\date{\today}
\begin{abstract}
The effect of the strong $\bar{\rm p}$-p nuclear interaction in a three-charge-particle
system with arbitrary masses is investigated. Specifically, the ($\bar{\rm p},\ \mu^-$,\ p)
system is considered, where $\bar{\mbox{p}}$ is an antiproton, $\mu^-$ is a muon and  p is a proton.
A numerical computation in the framework of a detailed few-body approach is carried out
for the following protonium (antiprotonic hydrogen) formation three-body reaction:
$\bar{\mbox{p}} + \mbox{H}_{\mu}(1s) \rightarrow (\bar{\mbox{p}} \mbox{p})_{\alpha} + \mu^-$.
Here, $\mbox{H}_{\mu}(1s)$ is a ground state muonic hydrogen, i.e. a bound state of p and $\mu^-$. 
A bound state of $\mbox{p}$ and its counterpart $\bar{\mbox{p}}$ 
is a protonium atom in a quantum atomic state $\alpha$, i.e. $Pn = (\bar{\mbox{p}}\mbox{p})_{\alpha}$.
The low-energy cross sections and rates of the $Pn$ formation reaction are computed in the framework of a
Faddeev-like equation. The strong $\bar{\rm p}$-p interaction is
included in these calculations within a first order approximation. It was found, that
even in the framework of this approximation the inclusion of the strong interaction
results in a quite significant correction to the rate of the three-body reaction. Therefore,
the title three-body antiprotonic process with participation of muons should be
useful, especially at low-energy collisions, in studying the $\bar{\rm p}$-p nuclear
forces and the annihilation channels in $Pn$.
\end{abstract}
\maketitle
\section{Introduction}
\label{sec:intro}
The first detection and exploration of antiprotons,
$\bar{\rm p}$'s, \cite{1st_pbar} occurred more than a half of a century ago.
Since that time this research field, which is related to stable baryonic particles,
has seen substantial developments in both experimental and theoretical aspects.
This field of particle physics represents one
of the most important sections of research work at CERN. It will suffice to mention such
experimental research groups as ALPHA \cite{andresen10}, ATRAP \cite{gabr11}, 
ASACUSA \cite{hori2013}  and others,
which carry out experiments with antiprotons.
By using slow antiprotons 
it is then possible to create ground state
antihydrogen atoms $\bar{\mbox H}_{1s}$ 
(a bound state of $\bar{\mbox p}$ and $e^+$, i.e. a positron) at low temperatures.
The resulting two-particle atom at present can be viewed as one 
of the simplest and most stable anti-matter species \cite{amoretti2002}.
A comparison of the properties of the resulting hydrogen atom H with $\bar{\mbox H}$ 
reveals that this antiatom lends itself well to support testing of the
fundamentals of physics \cite{nature2016}.
For example, an evaluation of the CPT theorem comes to mind immediately \cite{vargas2015}.
This possibility reinforces the need to obtain and store low-energy $\bar{\mbox p}$'s 
which could provide a basis for further comparisons of scientific interest. 
The system certainly contributes to the state of current research in both atomic and nuclear physics 
\cite{hori2013,madsen2015,gabr11,andresen10}.
Further, the basic idea could be expanded to other atoms. A good example of this would be metastable 
antiprotonic helium atoms (atomcules) such as
$\bar{\mbox{p}}^3$He$^+$ and $\bar{\mbox{p}}^4$He$^+$  \cite{hayano2007}.
It is important to note that within the field of $\bar{\mbox p}$ physics these Coulomb three-body systems are also
very important.
Specifically, the use of high-precision laser spectroscopy of atomclues allows one to measure
$\bar{\mbox{p}}$'s charge-to-mass ratio as well as fundamental constants within the standard model \cite{hori11}.
Developments in regard to atomcules and $\bar{\mbox H}$ atoms
have increased interest in the protonium ($Pn$) atom as well. This atom can be viewed as
a bound state of $\bar{\mbox p}$ and p \cite{zurlo2006,venturelli07,rizzini2012}.
The two-heavy-charge-particle system can also be described as antiprotonic hydrogen.
Its characteristics within the atomic scale are that it
is a heavy and an extremely small system containing strong Coulomb and nuclear interactions.
There is an interplay between these interactions inside the atom.
This situation is responsible for the creation of interesting
resonance and quasi-bound states in $Pn$ \cite{shapiro1}.
Thus, $Pn$ can be considered as a useful tool in the examination of the
antinucleon$-$nucleon $(\bar{N}N)$ interaction
potential \cite{np2015,jmr1992,jmr1982a,jmr1982} as well as the annihilation 
processes \cite{desai1961,jmr2002,jmr2005}.
In other words, the interplay between Coulomb and nuclear forces
contributes greatly to $\bar{\mbox p}$ and p quantum dynamics \cite{shapiro2}.
Further, the $\bar{\mbox p}+$p elastic scattering problem has also been examined in numerous papers. 
A good representative example would be paper \cite{jmr2002}. It is also worthwhile to note that $Pn$
formation is related to charmonium - a hydrogen-like atom $(\bar{c}c)$, which is also known as
a bound state of a $c$-antiquark $(\bar{c})$ and $c$-quark \cite{jmr2002}.
In sum, the fundamental importance of protonium and problems 
related to its formation, i.e. bound or quasi-bound states, 
resonances and spectroscopy, have resulted that this two-particle
atom gained much attention in the last decades.

Several few-charge-particle collisions can be used in order to produce low-energy $Pn$ atoms.
The following reaction is, for instance, one of them:
\begin{equation}
\bar{\mbox{p}} + \mbox{H}(1s) \rightarrow (\bar{\mbox{p}}\mbox{p})_{\alpha'} + e^-.
\label{eq:1st}\end{equation}
This process is a Coulomb three-body collision which
was computed in a few works in which different methods and
techniques have been applied \cite{tong06,sakimoto13,esry03}.    
Because in this three-body process a heavy particle, i.e. a
proton, is transferred from one negative "center", $e^-$, to another, $\bar{\mbox{p}}$,
it would be difficult to apply a computational method based 
on an adiabatic (Born-Oppenheimer) approach \cite{born1927}.
Besides, experimentalists use another few-body reaction to produce $Pn$ atoms, i.e.
a collision between a slow $\bar{\mbox{p}}$ and a positively charged molecular hydrogen ion, i.e. H$_2^+$:
$\bar{\mbox{p}} + \mbox{H}_2^+ \rightarrow (\bar{\mbox{p}}\mbox{p})_{\alpha'} + \mbox{H}.$
Nonetheless, this paper is devoted to another three-body collision of the $Pn$ formation
reaction in which we compute the cross-section and rate of a
collision between $\bar{\mbox{p}}$ and a muonic hydrogen atom H$_{\mu}$, which is a bound state
of p and a negative muon:
\begin{equation}
\bar{\mbox{p}} + (\mbox{p}\mu)_{1s} \rightarrow (\bar{\mbox{p}}\mbox{p})_{\alpha} + \mu^-,
\label{eq:3}\end{equation}
where, $\alpha$=$1s$, $2s$ or $2p$ is the final quantum atomic state of $Pn$.
Since the participation of $\mu^-$ in (\ref{eq:3}), at low-energy collisions $Pn$ would be formed in a very
small size - in the ground and close to ground states $\alpha$. It is obvious that
in these states the hadronic nuclear force between $\bar{\mbox{p}}$ and p will be strong and pronounced.
In its ground state the $Pn$ atom has the following size: $a_0(Pn)=\hbar^2/(e_0^2 m_p/2)\sim$ 50 fm,
in which the Coulomb interaction between $\bar{\mbox{p}}$ and p becomes extremely strong. 
The corresponding $Pn$'s binding energy without the inclusion of the nuclear $\bar{\rm p}$-p
interaction is: $E_n(Pn) = - e_0^4 m_p/2 / (2 \hbar n^2) \sim -$ 10 keV. We take:
$n=1$, $\hbar$ is the Planck constant, $e_0$ is the electron charge, and $m_p$ is the proton mass.
It would be useful to note, that the
realistic  $\bar{\rm p}$-p binding energy (with the inclusion of the
strong nuclear interaction) can have a large value. This value may be comparable or even larger than $m_p$. 
Consequently, it might be necessary to apply a relativistic treatment
to the reaction (\ref{eq:3}) in the output channel \cite{ueda}.
The situation which involves a very strong Coulomb interaction inside $Pn$
can also be a reason for vacuum polarization forces as well.
Therefore, within the reaction (\ref{eq:3}) it might be quite useful to take into account all these
physics effects and carry out a computation of their influence on the reaction's partial
cross sections and rates. Moreover, if in the near future it would be possible
to undertake a high quality measurement of 
(\ref{eq:3}), we could compare the new results with corresponding theoretical
data and fit (adjust) the $\bar{\rm p}$-p strong interaction into the theoretical
calculation in order to reproduce the laboratory data. This process will be useful in order to better understand  
the annihilation processes and the nature of the strong $\bar{\rm p}$-p interaction.
Muons are already used as an effective tool to search for
"new physics" and to carry out precise measurements of some fundamental constants \cite{lauss2009}.
For example, in the atomic analog of the reaction (\ref{eq:3}) $Pn$ would be formed at highly excited Rydberg
states with $\alpha^{\prime} \approx 30$.
Therefore, it is interesting to investigate the $\bar{\rm p}$-p nuclear interaction
in the framework of the three-body reaction (\ref{eq:3}) at low-energy collisions.
In this paper the reaction (\ref{eq:3}) is treated as a Coulomb
three-body system (123) with arbitrary masses: $m_1,\ m_2$, and $m_3$.
This is shown in Figs. \ref{fig1} and \ref{fig2}.
A few-body method based on a Faddeev-type equation formalism
is used. In this approach the three-body wave function is decomposed in two independent
Faddeev-type components \cite{fadd,fadd93}.
Each component is determined by its own independent Jacobi coordinates.
Since, the reaction (\ref{eq:3}) is considered at low energies, i.e. well below the three-body break-up threshold,
the Faddeev-type components are quadratically integrable
over the internal target variables $\vec r_{23}$ and $\vec r_{13}$. They are also shown in Fig. \ref{fig1}.
In this work the nuclear $\bar{\rm p}$-p interaction is included approximately by shifting the 
Coulomb (atomic) energy levels in $Pn$.
In the next sections we will introduce notations pertinent to the few-body
system (123), the basic equations, boundary conditions, and a brief derivation of the set of
coupled one-dimensional integral-differential equations. 
The muonic atomic units (m.a.u. or m.u.) are used in this work: $e = \hbar = m_\mu =1$,
$m_{\mu}=206.769\ m_e$ is the mass of the muon, $m_e$ is the electron mass,
the proton (anti-proton) mass is  $m_p = m_{\bar{p}}$=1836.152 $m_e$.

\section{A few-body approach}
The main thurst of this paper is
the three-body reaction (\ref{eq:3}).    
As we have already mentioned, a quantum-mechanical Faddeev-type few-body method is applied
in this work. A coordinate space representation is used. 
In general, the Faddeev approach is based on a reduction of the total
three-body wave function $\Psi$ on three Faddeev-type components \cite{fadd93}. However,
when one has two negative and one positive charges only two asymptotic
configurations are possible below the system's total energy $(E)$ break-up threshold.
This situation is explained in Fig. \ref{fig1} specifically for the case of the three-body system:
$\bar{\mbox{p}},\  \mu^-$ and $\mbox{p}^+$.
In the framework of an adiabatic hyperspherical close-coupling approach
the Coulomb three-body system has been considered in Ref. \cite{igarashi2008}.
Nevertheless, one can also apply a few-body type method to the three-body system in which
one can decompose $\Psi$ on two components and devise a set of two
coupled equations \cite{hahn72}. Additionally, it would be interesting to
investigate and estimate the effect of
the strong $\bar{\rm p}$-p nuclear interaction in the final state of the reaction (\ref{eq:3}).
This is done in the current work.
For a number of reasons the direct $\bar{\rm p}$-p annihilation channel in (\ref{eq:3})
is not included in the current calculations. This approximation is discussed at the end of the
following subsection.

\subsection{Coupled integral-differential equations}   \label{sec:fbd1}
A modified close coupling approach (MCCA) is applied in
this work in order to solve the Faddeev-Hahn-type (FH-type) equations 
\cite{my2000,my2013,my03}. In other
words, we carry out an expansion of the Faddeev-type components
into eigenfunctions of the subsystem Hamiltonians. This technique provides
an infinite set of coupled one-dimensional integral-differential equations.
Within this formalism the asymptotic of the full three-body wave function contains two
parts corresponding to two open channels \cite{merkur}.
One can use the following system of units: $e=\hbar=m_2=1$. 
We denote an antiproton $\bar{\rm{p}}$ by 1, a negative muon $\mu^-$  by 2, and a proton p by 3.
The total Hamiltonian of the three-body system is:
\begin{equation}
\hat{H} = \hat{H}_0 + V_{12}(\vec r_{12}) + V_{23}(\vec r_{23}) + \mathbb{V}_{13}(\vec r_{13}),
\end{equation}
where $ \hat{H}_0$ is the total kinetic energy operator of the three-body system,
$V_{12}(\vec r_{12})$ and $V_{23}(\vec r_{23})$ are Coulomb pair-interaction
potentials between particles 12 and 23 respectively, and:
\begin{equation}\label{eq:vcnn}
\mathbb{V}_{13}(\vec r_{13}) = V_{13}(\vec r_{13}) + v_{13}^{\bar{N}N}(\vec r_{13})
\end{equation}
is the Coulomb+nuclear interaction between particles 13, i.e. $\bar{\rm{p}}$ and p.
$v_{13}^{\bar{N}N}(\vec r_{13})$
is the $\bar{N}N$ strong short-range interaction between the particles. 
The last potential is considered as an approximate spherical symmetric intgeraction in this work.
The system is depicted in Figs. \ref{fig1} and \ref{fig2}
together with the Jacobi coordinates $\{\vec r_{j3}, \vec \rho_k\}$ and the different geometrical angles
between the vectors:
\begin{align}
&\vec r_{j3} = \vec r_3 - \vec r_j,\label{eq:coord1}\\
&\vec \rho_k = \frac{(m_3\vec r_3 + m_j\vec r_j)}{(m_3 + m_j)} - \vec r_k,\ \ \ (j \not = k=1, 2).\label{eq:coord2}
\end{align}
Here $\vec r_{\xi}$, $m_{\xi}$ are the coordinates and the
masses of the particles $\xi=1, 2, 3$ respectively.
This circumstance suggests a few-body Faddeev formulation which uses only two components.
A general procedure to derive such formulations is described in Ref. \cite{hahn72}.
In this approach the three-body wave function is represented as follows:
\begin{equation}\label{eq:psi3b}
|\Psi\rangle =  \Psi_1 (\vec r_{23},\vec \rho_1) + \Psi_2 (\vec r_{13},\vec \rho_2),
\end{equation}
where each Faddeev-type component is determined by its own Jacobi coordinates. Moreover,
$ \Psi_1 (\vec r_{23}, \vec \rho_1)$ is quadratically integrable
over the variable $\vec r_{23}$, and $\Psi_2 (\vec r_{13},\vec
\rho_2)$ over the variable $\vec r_{13}$. To define $|\Psi_l\rangle$, $(l = 1, 2)$ a set of two coupled
Faddeev-Hahn-type equations would be: 
\begin{align}
&\Big (E-\hat{H}_0-V_{23}(\vec r_{23}) \Big ) \Psi_1 (\vec r_{23}, \vec \rho_1) =
\Big (V_{23}(\vec r_{23}) + V_{12}(\vec r_{12})
\Big )\Psi_2 (\vec r_{13}, \vec \rho_2),\label{eq:fh1}\\
&\Big (E-\hat{H}_0 - \mathbb{V}_{13}(\vec r_{13})   
\Big )\Psi_2 (\vec r_{13}, \vec \rho_2) =
\Big (\mathbb{V}_{13}(\vec r_{13})       
+V_{12}(\vec r_{12}) \Big )\Psi_1 (\vec r_{23}, \vec \rho_1).\label{eq:fh2}
\end{align}
Here, $\hat{H}_0$ is the kinetic energy operator of the three-particle system, $V_{ij} (r_{ij})$
are paired Coulomb interaction potentials $(i \not= j = 1,2,3)$, $E$ is the total energy, and
$\mathbb{V}_{13}(\vec r_{13})$ is represented in Eq. (\ref{eq:vcnn}).
It is important to point out here, that
the constructed equations satisfy the Schr\H{o}dinger
equation exactly \cite{hahn72}. For the energies below the three-body
break-up threshold these equations exhibit the same advantages as the
Faddeev equations \cite{fadd}, because they are formulated for the wave
function components with correct physical asymptotes.

Next, the kinetic energy operator $\hat{H}_0$ in Eqs. (\ref{eq:fh1})-(\ref{eq:fh2})
can be represented as: $\hat{H}_0 = \hat{T}_{\rho_i} + \hat{T}_{r_{ij}}$, then
one can re-write the equations (\ref{eq:fh1})-(\ref{eq:fh2}) in the following way:
\begin{align}
& \Big (E-\hat{T}_{\rho_1}-\hat{h}_{23}(\vec r_{23})
\Big ) \Psi_1 (\vec r_{23}, \vec \rho_1) =
\Big (V_{23}(\vec r_{23}) + V_{12}(\vec r_{12})
\Big )\Psi_2 (\vec r_{13}, \vec \rho_2),\label{eq:fh11}\\
& \Big (E-\hat{T}_{\rho_2}-\hat{h}^{\bar{N}N}_{13}(\vec r_{13})
\Big )\Psi_2 (\vec r_{13}, \vec \rho_2) =
\Big (V_{13} (\vec r_{13}) + v^{\bar{N}N}_{13}(\vec r_{13})
+V_{12}(\vec r_{12})
\Big )\Psi_1 (\vec r_{23}, \vec \rho_1).\label{eq:fh22}
\end{align}
The two-body target hamiltonians $\hat{h}_{23}(\vec r_{23}) =  \hat{T}_{\vec r_{23}}+V_{23}(\vec r_{23})$ 
and $\hat{h}^{\bar{N}N}_{13}(\vec r_{13}) = \hat{T}_{\vec r_{13}}+V_{13}(\vec r_{13}) + v^{\bar{N}N}_{13}(\vec r_{13})$ 
with an additional $\bar{\rm p}$-p
nuclear interaction are represented explicitly in these equations. 
In order to solve Eqs. (\ref{eq:fh11})-(\ref{eq:fh22})
a modified close-coupling approach is applied, which leads to an expansion
of the system's wave function components $\Psi_1$ and $\Psi_2$
into eigenfunctions $\varphi^{(1)}_{n}(\vec r_{23})$ and $\varphi^{(2)\bar{N}N}_{n'}(\vec r_{13})$
of the subsystem (target) Hamiltonians, i.e.
\begin{eqnarray}
\left\{
\begin{array}{l}
\Psi_1(\vec r_{23}, \vec \rho_1)\approx\ \ \ {\large{{\mathclap{\displaystyle\int}\mathclap{\sum_{n}}}}}
\hspace{4mm}f^{(1)}_n(\vec \rho_1)\varphi^{(1)}_n(\vec r_{23}),\label{eq:fh77ras}\\
\Psi_2(\vec r_{13}, \vec \rho_2)\approx\ \ \ {\large{{\mathclap{\displaystyle\int}\mathclap{\sum_{n'}}}}}
\hspace{4mm}f^{(2)}_{n'}(\vec \rho_2)\varphi^{(2)\bar{N}N}_{n'}(\vec r_{13}).
\end{array}\right.
\end{eqnarray}
This provides a set of coupled one-dimensional integral-differential
equations after the partial-wave projection. 

The two complete sets of functions, i.e.
$\{\varphi_n^{(1)}(\vec r_{23})\}$ and $\{\varphi_{n'}^{(2)\bar{N}N}(\vec r_{13})\}$,
represent the eigenfunctions of the two-body target
hamiltonians $\hat{h}_{23}(\vec r_{23})$ and $\hat{h}^{\bar{N}N}_{13}(\vec r_{13})$ respectively:
\begin{align}
& \hat{h}_{23}(\vec r_{23})\varphi_n^{(1)}(\vec r_{23}) = 
\Big [\hat{T}_{\vec r_{23}}+V_{23}(\vec r_{23})\Big ]
\varphi_n^{(1)}(\vec r_{23}) = \varepsilon_n \varphi_n^{(1)}(\vec r_{23})\label{eq:h1}\\
& \hat{h}^{\bar{N}N}_{13}(\vec r_{13})\varphi^{(2)\bar{N}N}_{n'}(\vec r_{13}) =
\Big [\hat{T}_{\vec r_{13}}+V_{13}(\vec r_{13}) + v^{\bar{N}N}_{13}(\vec r_{13})\Big ]
\varphi^{(2)\bar{N}N}_{n'}(\vec r_{13}) = \mathcal{E}_{n'} 
\varphi^{(2)\bar{N}N}_{n'}(\vec r_{13})\label{eq:h2}
\end{align}
In addition to the Coulomb potential,    
the strong interaction, $v^{\bar{N}N}_{13}(\vec r_{13})$, is also
included in Eq. (\ref{eq:h2}).
Coulomb is a central symmetric potential. Therefore, the
eigenfunctions $\varphi_n^{(1)}$ and the corresponding eigenstates are \cite{landau}:
\begin{align}
&\varphi_{n}^{(1)}(\vec r_{23}) = \sum_{lm} R^{(1)}_{nl}(r_{23})Y_{lm}(\vec r_{23}),\label{eq:phi1} \\
&\varepsilon_n = -\frac{\mu_1}{2n^2}. \label{eq:epsil1}
\end{align}   

The full potential between $\bar{\rm p}$ and p is more complex,
because its second part, $ v^{\bar{N}N}_{13}(\vec r_{13})$,
posses an asymmetric $\bar{N}N$ nuclear interaction \cite{jmr2002,jmr2005}. 
We did not explicitly include the strong interaction in the current calculations.
Therefore, in the case of the target
$Pn$ eigenfunctions we used the two-body pure Coulomb (atomic) wave functions.
Nonetheless, the strong $\bar{\rm p}$-p interaction is approximately taken into account
in this work through the eigenstates $\mathcal{E}_{n'}$ which have
shifted values from the original Coulomb levels $\varepsilon_{n'}$ \cite{deser}, that is:
\begin{align}
&\varphi^{(2)\bar{N}N}_{n'}(\vec r_{13}) \approx \sum_{l'm'} 
R^{(2)\bar{N}N}_{n'l'}(r_{13})Y_{l'm'}(\vec r_{13})
\approx \sum_{l'm'} R^{(2)}_{n'l'}(r_{13})Y_{l'm'}(\vec r_{13})
\label{eq:phi2}\\
&\mathcal{E}_{n'} \approx \varepsilon_{n'} + \Delta E_{n'}^{\bar{N}N}=
-\frac{\mu_2}{2n'^2}+ \Delta E_{n'}^{\bar{N}N}.\label{eq:epsil2}
\end{align}
In Eqs. (\ref{eq:phi1}) and (\ref{eq:phi2}) $Y_{lm}(\vec r)$ are spherical functions \cite{varshal}
and $R^{(i)}_{nl}(r)$ $(i=1,2)$ is an analytical solution to the radial part of the two-charge-particle
Schr\H{o}dinger equation \cite{landau}:
\begin{equation}\label{eq:eqrnl}
\Big(\varepsilon_n^{(i)} + \frac{1}{2\mu_j r_{j3}^2}
\Big\{\frac{\partial}{\partial r_{j3}} \Big( r_{j3}^2 \frac{\partial}
{\partial r_{j3}} \Big) - l(l+1)\Big\} - V_{j3}\Big)R_{nl}^{(i)}(r_{j3})= 0,
\end{equation}
where $j \ne i =1, 2$. The method outlined above is only a first order approximation.
In the framework of this approach it would be interesting to estimate the level of influence
of the strong $\bar{\rm{p}}-$p interaction on the three-charge-particle proton transfer reaction (\ref{eq:3}).

Broadly speaking, the two-body Coulomb-nuclear wave functions of $Pn$, i.e. 
$\varphi^{(2)\bar{N}N}_{n'}(\vec r_{13})$ and corresponding eigenstates, $\mathcal{E}_{n'}$,
have been of a significant interest for a long time.
To build these states one needs to solve the two-charge-particle Schr\H{o}dinger
equation with an additional strong short-range $\bar{N}N$ interaction, i.e. Eq. (\ref{eq:h2}),
see for instance \cite{jmr1982}.
In Ref. \cite{armour2005} the authors explicitly included the nuclear $\bar{\rm{p}}$-p interaction
in the framework of a variational approach for the case of the $\bar{\rm{H}}$+H scattering.
However, as a first step, one can also apply an approximate approach:
Eqs. (\ref{eq:phi1})-(\ref{eq:phi2}) with an energy shift in the eigenstate of $Pn$ $\mathcal{E}_{n'}$, i.e.
Eq. (\ref{eq:epsil2}), $\varepsilon_{n'}$ is the Coulomb level and $\Delta E_{n'}^{\bar{N}N}$
is its nuclear shift. It can be computed, for example, with the use of the following
formula \cite{deser}:
\begin{equation}\label{eq:deser1}
\Delta E^{\bar{N}N}_{n'} = - \frac{4}{n'}\frac{a_s}{B_{Pn}}\varepsilon_{n'},
\end{equation}
where $a_s$ is the strong interaction
scattering length in the $\bar{\mbox{p}}+$p collision, i.e. without inclusion
of the Coulomb interaction between the particles, $B_{Pn}$ is the Bohr radius of $Pn$.
In the literature one can
find other approximate expressions to compute $\Delta E^{\bar{N}N}_{n'}$, 
see for example \cite{trueman,popov1979}.
It would also be interesting to apply some of these formulas   
in conjunction with the 
relativistic effects in protonium, see for example works \cite{ueda,thaler1983}.

After determining a proper angular momentum expansion one can obtain an infinite set of
coupled integral-differential equations for the unknown functions $f_{\alpha}^{(1)}(\rho_1)$ and
$f_{\alpha^\prime}^{(2)}(\rho_2)$ \cite{my2013}:
\begin{equation}  
\left \{
\begin{aligned}\label{eq:fh7}
\left[(k^{(1)}_n)^2\ +\ \frac{\partial^2}
{\partial \rho_1^2}\ -\
\frac{\lambda (\lambda + 1)}{\rho_1^2}\right]
f_{\alpha}^{(1)}(\rho_1) = g_1\sum_{\alpha'}
\frac{\sqrt{(2\lambda + 1)(2\lambda^{\prime} + 1)}}{2L+1}\\                                   
\times \int_{0}^{\infty} d \rho_{2}
f_{\alpha^\prime}^{(2)}(\rho_{2})\int_{0}^{\pi}
d \omega \sin\omega
R_{nl}^{(1)}(|\vec{r}_{23}|)
\left[-\frac{1}{|\vec {r}_{23}|} + \frac{1}{|\vec {r}_{12}|}\right]
R_{n'l'}^{(2)}(|\vec{r}_{13}|)\\                                                                                  
\times \rho_1 \rho_2 \sum_{mm'} D_{mm'}^L(0, \omega, 0)C_{\lambda 0lm}^{Lm}
C_{\lambda' 0l'm'}^{Lm'}
Y_{lm}(\nu_1, \pi) Y^*_{l'm'}(\nu_{2}, \pi),                                                              
\\ \\
\left[(k^{(2)}_n)^2\ +\ \frac{\partial^2}
{\partial \rho_2^2}\ -\
\frac{\lambda' (\lambda' + 1)}{\rho_2^2}\right] 
f_{\alpha}^{(2)}(\rho_2) =
g_2\sum_{\alpha'}
\frac{\sqrt{(2\lambda + 1)(2\lambda^{\prime} + 1)}}{2L+1}\\                                      
\times \int_{0}^{\infty} d \rho_{1}
f_{\alpha^\prime}^{(1)}(\rho_{1})\int_{0}^{\pi}
d \omega \sin\omega
R_{nl}^{(2)}(|\vec{r}_{13}|)
\left[-\frac{1}{|\vec {r}_{13}|} + \frac{1}{|\vec {r}_{12}|}\right]
R_{n'l'}^{(1)}(|\vec{r}_{23}|)\\                                                                                       
\times \rho_{2} \rho_1 \sum_{mm'} D_{mm'}^L(0, \omega, 0)C_{\lambda 0lm}^{Lm}
C_{\lambda' 0l'm'}^{Lm'}
Y_{lm}(\nu_2, \pi) Y^*_{l'm'}(\nu_{1}, \pi).                                                                    
\end{aligned}
\right.
\end{equation}  
Here: $g_i=4\pi M_i/\gamma^{3}$ $(i=1, 2)$, $L$ is the total angular 
momentum of the three-body system,
$\alpha=(nl\lambda)$ are quantum numbers of a three-body state,
$k_n^{(i)}=\sqrt{2M_i(E-E_n^{(j)})}$,
with  
$M_1 = (m_2+m_3)m_1/(m_1+m_2+m_3)$,
$M_2 = (m_1+m_3)m_2/(m_1+m_2+m_3)$,
$E_n^{(j)}$ is the binding energy of $(j3)$, $(i \ne j = 1, 2)$,
%
$\gamma=1-m_1m_2 / ( (m_1+m_3) (m_2+m_3) )$,
$D_{mm'}^L(0, \omega, 0)$ is the Wigner function \cite{varshal},
$C_{\lambda 0lm}^{Lm}$ is the Clebsh-Gordon coefficient \cite{varshal},
$\omega$ is the angle between the Jacobi coordinates
$\vec \rho_i$ and $\vec \rho_{i'}$, $\nu_i$ is the angle between
$\vec r_{i'3}$ and $\vec \rho_i$, $\nu_{i'}$ is the angle
between $\vec r_{i3}$ and $\vec \rho_{i'}$. The following relationships should be used
for the numerical calculations:
\begin{align}
& \sin \nu_i = \frac{\rho_{i'}}{\gamma r_{i'3}}  \sin\omega,\label{eq:most2}\\
& \cos \nu_i = \frac{1}{\gamma r_{i'3}}(\beta_i \rho_i + \rho_{i'} \cos \omega),\ \ \ (i \ne i' = 1,2).\label{eq:most3}
\end{align}

A detailed few-body treatment of the heavy-charge-particle reaction (\ref{eq:3})
is the main goal of this work. The geometric angles of the
configurational triangle $\triangle$123: $\nu_{1(2)}$, $\eta_{1(2)}$, $\zeta$, and $\omega$ are
shown in Fig. {\ref{fig2} together with the Jacobi coordinates, i.e. $\{\vec r_{j3},\ \vec \rho_{k}\}$ $(j \ne k = 1, 2)$
and $\vec r_{12}$. The center of mass of the (123) system is $O$. $O_1$ and $O_2$ are the center
of masses of the targets. The Faddeev decomposition avoids over-completeness problems because
the subsystems are treated in an equivalent way in the framework of the two-coupled equations.
Thus, the correct asymptotes are guaranteed. The Faddeev-components are smoother functions of
the coordinates than the total wave function \cite{fadd93,merkur}.

In the framework of the first order approximation approach the direct $\bar{\rm p}$-p
annihilation channel in the reaction (\ref{eq:3}) is not included in this work.
In the input channel of the reaction (\ref{eq:3}), $\bar{\mbox{p}}$+(p$^+\mu^-)_{1s}$,
the relatively heavy muon very effectively screens the strong Coulomb potential of the proton,
and therefore it significantly prevents direct annihilation in (\ref{eq:3}) before the 
$Pn$ formation. In other words, the $Pn$ formation process dominates. However,
it is another matter in the case of the atomic version of the $Pn$ formation
reaction (\ref{eq:1st}). Here, the electron cloud around the proton can also block the 
$\bar{\mbox{p}}$ movement to p, but because of the quantum-tunneling effect the massive 
antiproton can penetrate with a significant probability
through the light electron cloud and then directly annihilate with proton before protonium forms.
Therefore, in the framework of the reaction (\ref{eq:1st}) it would be necessary to take into account
the tunneling effect. As far as we know, this is still not done in a suitable way.

In terms of the $Pn$ annihilation in the reaction (\ref{eq:3}) (which can occur 
after the two-body system formation) and an inclusion of this effect
in calculations, it was mentioned above that in this case one needs to build precise
Coulomb-nuclear $\bar{\rm p}$-p
two-body wave functions $\varphi^{(2)\bar{N}N}_{n'}(\vec r_{13})$ from Eq. (\ref{eq:h2}).
In this special case, one needs to consider not only
the shifts of the Coulomb levels Eqs. (\ref{eq:epsil2}), but also their widths.
However, in the current work, as a first order approximation the nuclear effect is considered
only through Eqs. (\ref{eq:epsil2}) and (\ref{eq:deser1}).

We believe that to some extent this approximation is justified. In this work, we were mostly interested
in the $Pn$ atom formation process (\ref{eq:3}),
where the values of the Coulomb-nuclear atomic levels at which
the atom can form are important. As we mentioned, these levels have widths, but they are
mostly responsible for the annihilation reaction that follows.

\subsection{Boundary conditions}  \label{subsec:cs}
To reach the next step it is necessary to obtain a unique solution for equations (\ref{eq:fh7}).
While doing so it is important that the appropriate boundary conditions are chosen.
They should be related to the physical situation of the system. The following condition is imposed first:
\begin{equation}\label{eq:bound0}
f_{nl}^{(i)}(0) \mathop{\mbox{\large$\sim$}}0.
\end{equation}
Subsequently, it is then appropriate to solve the three-body 
charge-transfer problem to utilize the ${\bf K}-$matrix formalism approach.
This would appear to be a prudent step because this method 
has been successfully used to obtain solutions in various three-body problems
within the framework of both the Schr\H{o}dinger equation \cite{melezhik,cohen91} and
the coordinate space Faddeev equation \cite{kvits1995}.
Specifically, in regard to the rearrangement scattering  problem $i +(j3)$
as the initial state within the asymptotic region it will be necessary to devise two solutions to
Eqs. (\ref{eq:fh7}) which then will satisfy the boundary conditions that follow:
\begin{eqnarray}
\left\{
\begin{array}{l}
f_{1s}^{(i)}(\rho_i)
\mathop{\mbox{\large$\sim$}}\limits_{\rho_i \rightarrow + \infty}
\sin(k^{(i)}_1\rho_i) + K_{ii}\cos(k^{(i)}_1\rho_i)\; 
\vspace{1mm}\\
f_{1s}^{(j)}(\rho_j)
\mathop{\mbox{\large$\sim$}}\limits_{\rho_j \rightarrow + \infty}
\sqrt{v_i / v_j}K_{ij}
\cos(k^{(j)}_1\rho_j)\;,\\
\end{array}\right.
\label{eq:fh5}
\end{eqnarray}
where $\it K_{ij}$ represents the appropriate scattering coefficients,
and $v_{i(j)}$ ($i \ne j = 1, 2$) is the $i(j)$ channel velocity between the particles.
Next, one can use the following change of variables in Eq. (\ref{eq:fh7}), i.e.
\begin{equation}\label{eq:replace}
{\sf f}_{1s}^{(i)}(\rho_i)=f_{1s}^{(i)}(\rho_i)-\sin(k^{(i)}_{1}\rho_i),\ \ \ (i=1, 2).
\end{equation}
This substitution results in a modification of the variables
and provides two sets of inhomogeneous equations which can now be
conveniently solved numerically. Some details of our numerical approach
are presented below (See Appendix).
The transition also allows the coefficients $K_{ij}$ to be gained by reaching a
numerical solution for the previously described FH-type equations.
Now the cross section can be expressed as follows:
\begin{equation}\label{eq:crosssec}
\sigma_{ij} =
 \frac{4\pi}{k_1^{(i)2}}\left |\frac{{\bf K}}
{1 - i{\bf K}}\right |^2 = \frac{4\pi}{k_1^{(i)2}}\frac{\delta_{ij}D^2 + {\it K}_{ij}^2}
{(D - 1)^2 + ({\it K}_{11} + {\it K}_{22})^2}\ ,
\end{equation}
where ($i,j=1,2$) refer to the two channels and
$D = K_{11} K_{22} - K_{12} K_{21}$.
Next, in accord with the quantum-mechanical unitarity principle the
scattering matrix $\bf K = \begin{pmatrix} K_{11} & K_{12} \\ K_{21} & K_{22} \end{pmatrix}$
has an important feature, i.e. $K_{12}=K_{21}$, or:   
\begin{equation}
\chi (E) = \frac{K_{12}}{K_{21}} = 1.
\label{eq:unitarity}
\end{equation}
The last equation   
has been checked for all considered collision
energies within the framework of the 1s, 1s+2s and 1s+2s+2p 
MCCA approximations, i.e. Eqs. (\ref{eq:fh77ras}).

\section{Results}
\label{sec:results}
In this section we present our results. The $Pn$ formation three-body reaction is computed at low energies.
A Faddeev-like equation formalism Eqs. (\ref{eq:fh11})-(\ref{eq:fh22})
has been applied. The few-body approach has been explained in previous sections. 
In order to solve the coupled equations
two different independent sets of target expansion functions have been employed (\ref{eq:fh77ras}).
In the framework of this approach the two targets are treated equivalently and
the method allows us to avoid the over-completeness problem.
The goal of this paper is to carry out a reliable quantum-mechanical computation of the cross 
sections and corresponding rates of the $Pn$ formation reaction at low and very low collision energies.
It is very interesting to estimate the influence of the strong short-range $\bar{\rm p}$-p interaction
on the rate of the reaction (\ref{eq:3}). The three-body reaction
(\ref{eq:3}) could be used to investigate the strong $\bar{\rm p}$-p nuclear interaction and the
annihilation process in future experiments with the anti-protonic hydrogen atom or protonium $Pn$.
The coupled integral-differential Eqs. (\ref{eq:fh7}) have been solved numerically for the case of
the total angular momentum $L=0$ in the framework of the two-level  2$\times$(1s), four-level 2$\times$(1s+2s), 
and six-level 2$\times$(1s+2s+2p) close coupling approximations in Eq. (\ref{eq:fh77ras}). The sign "2$\times$"
indicates that two different sets of expansion functions are applied. The $L=0$ computation is justified,
because we are interested in a very low-energy collision: $\varepsilon_{coll}\sim10^{-4}$ eV$-$10 eV.
The following boundary conditions (\ref{eq:bound0}), (\ref{eq:fh5}), and (\ref{eq:replace}) have been applied.
To compute the charge transfer cross sections the expression (\ref{eq:crosssec}) has been used.

Below we report the computational results. However, before attempting large scale production calculations one needs
to investigate numerical convergence of the method and the computer program.
Fig. \ref{fig6} depicts a few of the initial convergence results for the case of the $1s+2s$ MCCA approach.
Specifically, in this case we solved four coupled integral-differential equations. 
The polarization effect, however, is not included.
In Fig. \ref{fig6} one can see,
that the inclusion of only the short-range $s$-states in the expansion (\ref{eq:fh77ras})
provides stable results for the rate, $\sigma_{tr}v_{c.m.}$ (upper plot), 
and for the transfer cross section, $\sigma_{tr}$ (middle plot).
Here, $v_{c.m.}= \sqrt{2\varepsilon_{coll}/M_k}$ is a relative center-of-mass 
velocity between the particles in the input channel of the three-body
reaction, $\varepsilon_{coll}$ is the collision energy, and $M_k$ is the reduced mass.
The upper limit of the integration can be taken as $R\approx$13 m.a.u. or 20 m.a.u.
A large number of integration points was used and we obtained a fully convergent result. 

Because we compared the $Pn$ formation 
rates, $\sigma_{tr}v_{c.m.}$,  
of the process (\ref{eq:3})
with the corresponding results from Ref. \cite{igarashi2008}, 
we also multiplied our data by factor of "$\times 5$",
as was done in \cite{igarashi2008}.   
Next, the COND number (Fig. \ref{fig6}, lower plot)
is an important special parameter of the DECOMP computer
program from \cite{forsythe}. The program is included and used in our FORTRAN code.
DECOMP solves the large system of linear equations (\ref{eq:l7}).
COND shows the quality of the numerical solution of a large system
of linear equations \cite{forsythe}. One can see that COND maintains quite constant values,
when energy changes from 10$^{-4}$ eV to 10 eV. It shows that our calculations are quiet stable.
However, COND increases its values when the upper limit of integration is increased.

Figs. \ref{fig7} and \ref{fig8} represent the convergence results in the framework of the
$1s+2s+2p$ MCCA approach in which we solve six coupled integral-differential equations.
In these cases we used a different number of
integration points, namely 75 and 85 per the muonic radius length and also varied the values of the upper
limit of the integration to 62, 69 and 76 m. a. u. Thus, the maximum number of integration knots
used in this work is $N_{max} = 76\times 85 = 6460$. It is seen that the results are in a good agreement
with each other in regard to the transfer and the elastic cross sections.
Thus, numerical convergence has been achieved.

We compared some of our findings with the corresponding data from the older work \cite{igarashi2008}.
The $Pn$ formation cross section in the reaction (\ref{eq:3}) are shown in Fig. \ref{fig9}.
Here we use $1s$, $1s+2s$ and $1s+2s+2p$ states within the modified close-coupling
approximation, i.e. MCCA approach. 
One can see that the contribution of the $2s$- and $2p$-states
from each target is becoming even more significant while the collision energy becomes smaller.
It is useful to make a comment about the behavior of $\sigma_{tr}(\varepsilon_{coll})$
at very low collision energies: $\varepsilon_{coll}\sim0$.
From our calculations we found the following relationship in the
p transfer cross sections: $\sigma_{tr}\rightarrow \infty$ 
as $\varepsilon_{coll}\rightarrow 0$. 
However, the p transfer rates, $\lambda_{tr}$, are proportional to the product
$\sigma_{tr} v_{c.m.}$ and this trends to a finite value as $v_{c.m.}\rightarrow 0$. 

To compute the proton transfer rate the following formula
$\lambda_{tr} = \sigma_{tr}(\varepsilon_{coll}\rightarrow 0)v_{c.m.}$  
can be used.
Therefore, additionally, for process (\ref{eq:3}) we can compute the numerical value of the following important quantity:
\begin{equation}
\Lambda(Pn) = \sigma_{tr}(\varepsilon_{coll} \rightarrow 0)v_{c.m.}\approx \mbox{const}, 
\end{equation}
which is proportional to the actual $Pn$ formation rate at low collision energies.
In the framework of the 2$\times(1s+2s+2p)$ MCCA approach, i.e. when six coupled Faddeev-Hahn-type
integral-differential equations are solved, our result for the $Pn$ formation rate has the following value:
\begin{equation}
\Lambda_{1s2s2p}(Pn) 
\approx 0.32\  \mbox{m.a.u}.
\end{equation}
The corresponding rate from work \cite{igarashi2008} is: $\Lambda' (Pn) \approx 0.2$ m.a.u. Both of these
results are in agreement with each other. For comparison purposes 
our original result for $\Lambda_{1s2s2p}(Pn)$ has been
multiplied by a factor of "$\times 5$" to match work \cite{igarashi2008}.

One of the main goals of this work is to investigate the effect of the $\bar{\rm p}$-p nuclear
interaction on the rate of the reaction (\ref{eq:3}).  
In Fig. \ref{fig9} we additionally provide our cross sections for (\ref{eq:3}) including the nuclear effect
in the final $Pn$ state. One can see, that the contribution
of the strong interaction becomes even more substantial when the collision energy becomes lower.
Also, for a few selected energies Table \ref{table1} shows our results
for the $Pn$ formation total cross sections and rates in the framework of different MCCA approximations.
The unitarity relationship, i.e. Eq. (\ref{eq:unitarity}), is checked. It is seen, 
that $\chi$ exhibits fairly constant values close to one.
A few additional comments about the inclusion of the $\bar{\rm p}$-p
nuclear interaction are appropriate.
First of all, we neglected the $\bar{\mbox p}$+p annihilation
channel. This approximation has been discussed above.
However, the effect of the strong nuclear forces on the reaction (\ref{eq:3})
is incorporated through the energy shifts $\Delta E_{n'}^{\bar{N}N}$ to the original Coulomb energy
levels in the $Pn$ atom, i.e. $\varepsilon_{n'}$ in Eq. (\ref{eq:epsil2}). To compute $\Delta E_{n'}^{\bar{N}N}$
the expression (\ref{eq:deser1}) is used from \cite{deser}. The $\bar{\mbox p}+$p elastic scattering length,
i.e parameter $a_s$, was adopted from
work \cite{jmr1992} and equals 0.57 fm in our calculations.
In \cite{jmr1992} the Kohno-Weise strong interaction potential \cite{KW1986} has been applied.  

The next two Figs. \ref{fig10} and \ref{fig11}
represent results in which we compare cross sections and rates computed with and without
the inclusion of the strong potential within the different close-coupling approximation.
Fig. \ref{fig10} shows our results in the framework of the $1s$ and $1s+2s$ MCCA approaches.
The results are numerically stable.
It seen that the contribution of the strong nuclear interaction is higher in the case of the $1s+2s$
approximation. For example, in this case the rate of
the reaction (\ref{eq:3}) is about 0.12 m.a.u., however with the inclusion of the nuclear interaction
it becomes 0.15 m.a.u. The last figure in this paper, Fig. \ref{fig11}, represents our computational data
in the $1s+2s+2p$ approach. The very important polarization effect is included. The inclusion of
the nuclear interaction brings a significant change to the rate of the reaction (\ref{eq:3}). At very
low collision energies around $10^{-4}-10^{-2}$ eV the rate is $\sim$0.5 m.a.u. It is important
to restate that all calculations carried out in this work have been done for the
ground-to-ground state of (\ref{eq:3}), i.e. $\alpha=1$.

\section{Conclusion}
In summation, the complexity of the few-body system and the method 
utilized necessitated that only the total orbital momentum $L=0$ be taken into account. 
However, the method was indeed adequate for
the slow and ultraslow collisions discussed previously.
Further, it is important to note that the devised few-body equations (\ref{eq:fh1})-(\ref{eq:fh2})
do exactly satisfy the Schr\H{o}dinger
equation. In cases in which the energies below the three-body
break-up threshold occur this methodology provides advantages similar to the
Faddeev equations \cite{fadd93}. 
This is because these equations are formulated to include wave
function components which contain the correct physical asymptotes.
The solution of these equations begins by using a close-coupling approach.
This then leads to an expansion
of the system's wave function components into
eigenfunctions of the subsystem (target) Hamiltonians, which
results in a set of one-dimensional integral-differential equations upon completion of
the partial-wave projection.

In an effort to expand the scope of the results a strong proton-antiproton
interaction was included by appropriately shifting the
Coulomb energy levels of the $Pn$ atom \cite{jmr1982,deser}.
Interestingly, this process increased the magnitude of the resulting values of the reaction
cross section and corresponding rate by $\sim 50$\%. Therefore, one further three-body
reaction similar to (\ref{eq:3}) can also be of a sufficient future interest:
\begin{equation}
\bar{\mbox{p}} + {^2}\mbox{H}_{\mu}(1s) \rightarrow (\bar{\mbox{p}} \mbox{d})_{\alpha} + \mu^-,
\label{eq:33}\end{equation}
where $^2$H=d is the deuterium nucleus, $\mu^-$ and $\bar{\mbox{p}}$ are muon and antiproton respectively.
This is because of a possible effect of the
isotopic few-body quantum dynamic differences between reactions (\ref{eq:3}) and (\ref{eq:33}),
and the nuclear interaction differences between $\bar{\mbox{p}}$ and p and $\bar{\mbox{p}}$ and d.
In the future, it would be very interesting to compare the cross sections of both reactions.

Based on the results herein it seems logical for future work to include in Eqs. (\ref{eq:fh77ras})
the higher atomic target states such as $3s+3p+3d+4s+4p...$ as well as the continuum spectrum.
Calculations of this type would be very interesting but challenging.
The challenge is because at very low energy collisions   
the higher energy channels are closed and there is 
a significant energy gap between the states and the actual collision energies.
Despite this limitation the primary contribution from s- and p-states (polarization)
is still evaluated. In closing, the authors feel that including the strong $\bar{\rm p}$-p
interaction explicitly in the numerical solution of Eqs. (\ref{eq:fh11})-(\ref{eq:fh22})
could also provide an interesting and challenging direction for future theoretical research in this area.

\begin{acknowledgements}
This paper was supported by the 
Office of Research and Sponsored Programs of St. Cloud State University, USA
and FAPESP and CNPq of Brazil.
\end{acknowledgements}


\appendix*
\section{Numerical method and solutions}
\label{numerics}
The delicacy of the three-charge-particle system consideration consists in the fact that
the Coulomb potential is a singular function. This singularity is a major
problem in numerical calculations involving few-body systems with Coulomb potentials.
Below we provide a brief discussion of our numerical approach used in this paper.
It would be somewhat simpler to reach a numerical solution for
the set of coupled Eqs. (\ref{eq:fh7}) if only the most important -s and -p waves
are included within the expansions (\ref{eq:fh77ras}), (\ref{eq:phi1}) and (\ref{eq:phi2}),
and limit $n$ up to $n=2$ in the Eq. (\ref{eq:fh77ras}).
This process results in a truncated 
set of six coupled integral-differential equations because in $\Psi_{1(2)}$ 
only 1s, 2s and 2p target two-body atomic wave-functions are included.
This method could be considered as a modified version
of the close coupling approximation containing six expansion functions.
The resulting set of truncated integral-differential Eqs. (\ref{eq:fh7})
may be solved by using a discretization procedure. Specifically,
on the right side of the equations the integrals over $\rho_1$ and $\rho_2$
can be replaced with sums using the trapezoidal rule \cite{abram}.
Further, the second order partial derivatives on the left side can be
discretized by using a three-point rule \cite{abram}. 
This process allows us to obtain a set of linear equations for the unknown coefficients
$f^{(i)}_{\alpha}(k)$ ($k=1, N_p$) \cite{my2013,my03}.
Then it is possible to ascertain through the symbolic-operator notations that the set of linear equations
has the following characteristics \cite{my2013,my03}:
\begin{equation}\label{eq:l7}
\sum_{\alpha'=1}^{2\times N_s}\sum_{j=1}^{N_p} {\bf A}_{\alpha \alpha'}(i,j) \vec f_{\alpha'}(j) =\vec b_{\alpha}(i).
\end{equation}
The resulting discretized equations can be then solved using the Gauss elimination method \cite{forsythe}.
It then follows that
the matrix $\bf A$ should exhibit a well known block-structure. In this case there are four major blocks 
in the matrix: two of them are related to the differential operators 
and other two are related to the integral operators \cite{my2013,my03}.
Further, each block should contain sub-blocks. The number of 
sub-blocks, of course, depends on the quantum numbers $\alpha=nl\lambda$ and
$\alpha' = n'l'\lambda'$. It is worth noting that the second order 
differential operators produce three-diagonal sub-matrixes \cite{my03}.

The solution to the coupled integral-differential equations (\ref{eq:fh7}) requires one to first compute 
the angular integrals $S_{\alpha \alpha'}^{(ii')}(\rho_i, \rho_{i'})$ $(i \ne i' = 1, 2)$ \cite{my2013}.
These integrals are independent of the energy, $E$.
To improve efficiency, one can compute them only once and then store them on a computer's hard drive for future reference. 
For example, they could be used in the calculations required to determine charge-transfer cross-sections at different collision energies. 
Also noteworthy is the relationship of the sub-integral expressions which
have a very strong and complicated dependence 
on the Jacobi coordinates $\rho_i$ and $\rho_{i'}$ \cite{my2013}. The next three figures presented \ref{fig3}, \ref{fig4} and
\ref{fig5} depict some of these relationships using different quantum numbers $\alpha$ and $\alpha'$.
Specifically, Fig. \ref{fig3} depicts the result in a case where $\alpha = \alpha' =1s$. In other words, this case illustrates a crucial
ground-state to ground-state matrix element in Eqs. (\ref{eq:fh7}) within the input channel. Further evaluation reveals
that this surface has smaller numerical values relative to the matrix element shown
in Fig. \ref{fig4}. In this case $\alpha = 1s$ and $\alpha' = 2p$, which assumes that the polarization effect is taken into account.
An interesting case when $\alpha = \alpha' =2s$ in the input channel is shown in Fig. \ref{fig5}.
Further, the analysis reveals that a very strong polarization effect results in the input channel of the reaction (\ref{eq:3}).
To clarify this further, one can calculate $S_{\alpha \alpha'}^{(ii')}(\rho_i, \rho_{i'})$ 
at different values of $\rho_i$ and $\rho_{i'}$. To do this an adaptable algorithm has been devised and applied using 
the following mathematical substitution \cite{my2013}:
$
\cos \omega = (x^2 - \beta_i^2\rho_i^2 - \rho_{i'}^2) / (2\beta_i \rho_i\rho_{i'}).
$
The angle dependent portion of the resulting equation can be described by the following one-dimensional integral:
\begin{eqnarray}\label{eq:omega}
S_{\alpha \alpha'}^{(ii')}(\rho_i, \rho_{i'}) =
\frac{4\pi}{\beta_i}
\frac{[(2\lambda + 1)(2\lambda^{\prime} + 1)]^{\frac{1}{2}}}{2L+1}
\int_{|\beta_i\rho_i - \rho_{i'}|}^{\beta_i\rho_i + \rho_{i'}}
dx R_{nl}^{(i)}(x) 
\left[-1 + \frac{x}{r_{ii'}(x)}\right] R_{n'l'}^{(i')}(r_{i3}(x))\nonumber \\
\times  \sum_{mm'} D_{mm'}^L(0, \omega(x), 0)C_{\lambda 0lm}^{Lm}
C_{\lambda' 0l'm'}^{Lm'} 
Y_{lm}(\nu_i(x), \pi) Y^{*}_{l'm'}(\nu_{i'}(x), \pi).\ \  
\end{eqnarray}

Specifically, the adaptive algorithm incorporated in a 
FORTRAN subroutine from \cite{berlizov99} is used within this work to calculate the angle integration in (\ref{eq:omega}).
This recursive computer program known as QUADREC,
is an improved version of the well respected program QUANC8 \cite{forsythe}. 
Therefore, QUADREC can provide improvements in regard to quality, 
stability and precision in integration when compared to QUANC8 \cite{berlizov99}.
When considering the expression (\ref{eq:omega})
it is worth noting that it differs from zero only in a quite narrow strip, i.e. when $\rho_i \approx \rho_{i'}$.
This can be explained because in the three-body system considered the coefficient $\beta_i$ approximately equals one. 
This means to it is imperative, to distribute a very large number 
of discretization points (up to 6000) between 0 and $\sim$80 muonic units if numerically reliable
converged results are to be reached.

We mentioned above, that
the truncated set of coupled integral-differential equations (\ref{eq:fh7}) is
solved with the use of the matrix approach. The
computation itself is organized in the following way: as a
first step two sets of  integration knots are created
over the Jacobi coordinates $\rho_1$ and $\rho_2$, i.e. we have:
$\{\rho_{1i}, i=1, N_1\}$ and $\{\rho_{2j}, j=1, N_2\}$. We  choose $N_1=N_2=N$,
where $N$ is taken up to 6500 points. Within the second step of the method
we have to carry out a numerical computation of the angle integrals (\ref{eq:omega}) for each
given coordinate value: $\rho_{1i}$ and $\rho_{2j}$. A special FORTRAN adaptive-quadrature
subroutine is used in this work.
Because of the very singular character of the Coulomb pair-interaction potentials between the particles
this step is  important but very challenging and time consuming.
Based on our observation, 
the precision and quality of these calculations should be robust enough.
The calculated angle integrals, Eq. (\ref{eq:omega}) can be saved on a hard drive of a computer system.
After this initial, but very important
work our program builds the full matrix which precisely corresponds to the set of coupled
Eqs. (\ref{eq:fh7}). Finally, one can solve the set of the linear equations (\ref{eq:l7}),
compute the three-body wave function, the elastic and charge-transfer cross sections,
and the corresponding $Pn$ formation rates.

Also, it would be useful to make few additional comments about the structure of the Eqs. (\ref{eq:fh7})
and our numerical method.  Namely, on the left side of these equations we have the usual differential operators.
However, because coupled Faddeev-Hahn-type equations are used in this work, on the right side of the equations
Eqs. (\ref{eq:fh7}) we have the unknown functions under the integration over $\rho_1$ and $\rho_2$.
The integration runs from 0 to infinity.
Therefore, it is obvious, that the usual step-by-step or predictor-corrector numerical 
methods in which a computer program itself (automatically) adopts integration steps \cite{forsythe}
cannot be applied in these calculations. Consequently, in the current case one needs, first, to choose and
distribute the integration knots in accord with the peculiarities of the potential surfaces (\ref{eq:omega}) 
and then build the full matrix. 
In turn the surfaces (matrix elements) $S_{\alpha \alpha'}^{(ii')}(\rho_i, \rho_{i'})$
have quite complicated and very different shapes and values.
This is seen in Figs. \ref{fig3}, \ref{fig4}, and \ref{fig5}. It is very important not
to lose all these peculiarities and carefully distribute as well as utilize a
large number of integration points. 
However, the last circumstance results in a very large matrix.

Another complication arises from the fact that the reaction (\ref{eq:3}) in the input channel has
a muonic atom H$_{\mu}$ as a target, but in the output channel $Pn$ is present. The size of
$Pn$ is about five times smaller than H$_{\mu}$, 
therefore the number of the integration knots which are sufficient
to describe the H$_{\mu}$ channel may not be good enough to compute the $Pn$ channel.

\newpage

\begin{center}
\Huge{TABLE I and FIGURES 1$-$11}
\end{center}

\clearpage
\begin{table}\label{tab1}
\caption{The total $Pn$ formation cross sections $\sigma_{tr}(\varepsilon_{coll})$ and rates $\Lambda(Pn)$
in the three-body reaction (\ref{eq:3}), when $\alpha=1$ and $\varepsilon_{coll}$ is the collision energy.
The results are presented in the framework of the different MCCA approach: $1s$, $1s+2s$, and $1s+2s+2p$.
The cross section $\sigma_{tr}$ is given in cm$^2$ and $\Lambda(Pn)$ in m.a.u. The unitarity condition
coefficient $\chi$, i.e. Eq. (\ref{eq:unitarity}), is also shown. Results with the inclusion of 
the strong nuclear interaction between $\bar{\mbox{p}}$ and p are presented only in the $1s+2s+2p$
approximation.  
For convenience, rates $\Lambda$'s have been multiplied 
by factor of "$\times 5$"  in this table, as in Ref. \cite{igarashi2008}. The rate 
with inclusion of the nuclear interaction, i.e. $\Lambda^{\bar{\rm{p}}\rm{p}}$,
is also multiplied by the same factor.}
\vspace{4mm}
\centering
\begin{tabular}{lccccccccc}
\hline
&\multicolumn{2}{c}{$1s$}
&\multicolumn{3}{c}{$1s+2s$}
&\multicolumn{2}{c}{$1s+2s+2p$}
&\multicolumn{2}{c}{$1s+2s+2p$, Nucl. $\bar{\rm{p}}$-$\rm{p}$}\\[3pt]
\hline 
%
$\varepsilon_{coll}$& $\sigma_{tr}$ & $\Lambda(Pn)$
                               & $\sigma_{tr}$ & $\Lambda(Pn)$ & $\chi$
                               & $\sigma_{tr}$ & $\Lambda(Pn)$
                               & $\sigma^{\bar{\rm{p}}\rm{p}}_{tr}$
                               & $\Lambda^{\bar{\rm{p}}\rm{p}}(Pn)$\\ 
%
\hline 
0.0001  & 1.3E-19  & 0.08639  & 1.9E-19  & 0.1269 & 0.97 & 4.95E-19  & 0.3251  &  7.55E-19  & 0.5027\\
0.001    & 4.1E-20  & 0.08639  & 6.0E-20  & 0.1269 & 0.98 & 1.57E-19  & 0.3249  &   2.39E-19 & 0.5025\\
0.05      & 5.8E-21  & 0.08636  & 8.5E-21  & 0.1269 & 0.97 & 2.18E-20  & 0.3193  &   3.33E-20 & 0.4950\\
1.0        & 1.3E-21  & 0.08593  & 1.9E-21  & 0.1273 & 0.97 & 2.89E-21  &  -            &   5.22E-21 & -           \\
10.0      & 3.9E-22  & 0.08183  & 6.4E-22  & 0.1343 & 0.97 &                  &  -           &  -                & -           \\[3pt]
\hline{\smallskip}\\
\end{tabular}\label{table1}
\end{table}

\clearpage
\begin{figure*}
\centering
\includegraphics[width=0.75\textwidth]{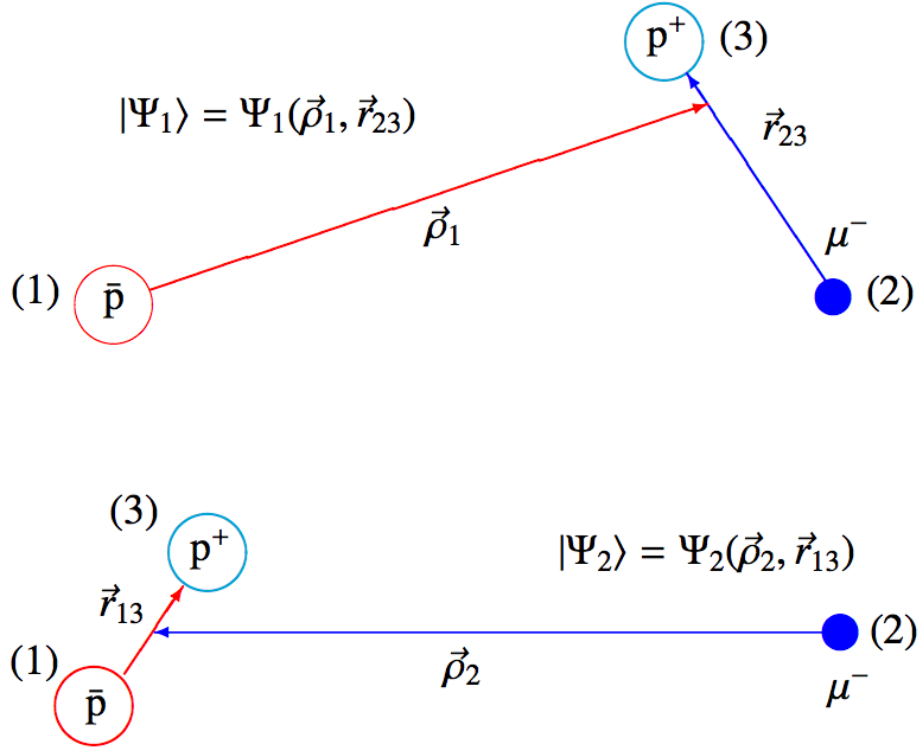}
\caption{Two asymptotic spacial configurations of the three-body system (123), or more specifically
$(\bar{\mbox{p}}, \mu^-, \rm{p}^+)$, which are considered in this work. The few-body
Jacobi coordinates $(\vec \rho_i, \vec r_{jk})$, where $i\ne j\ne k=1,2,3$  are also shown together with the 
three-body wave function
components $\Psi_1$ and $\Psi_2$: $\Psi = \Psi_1 + \Psi_2$ is the total wave function of the three-body system.}
\label{fig1}\end{figure*}

\clearpage
\begin{figure*}
\centering
\includegraphics[width=0.75\textwidth]{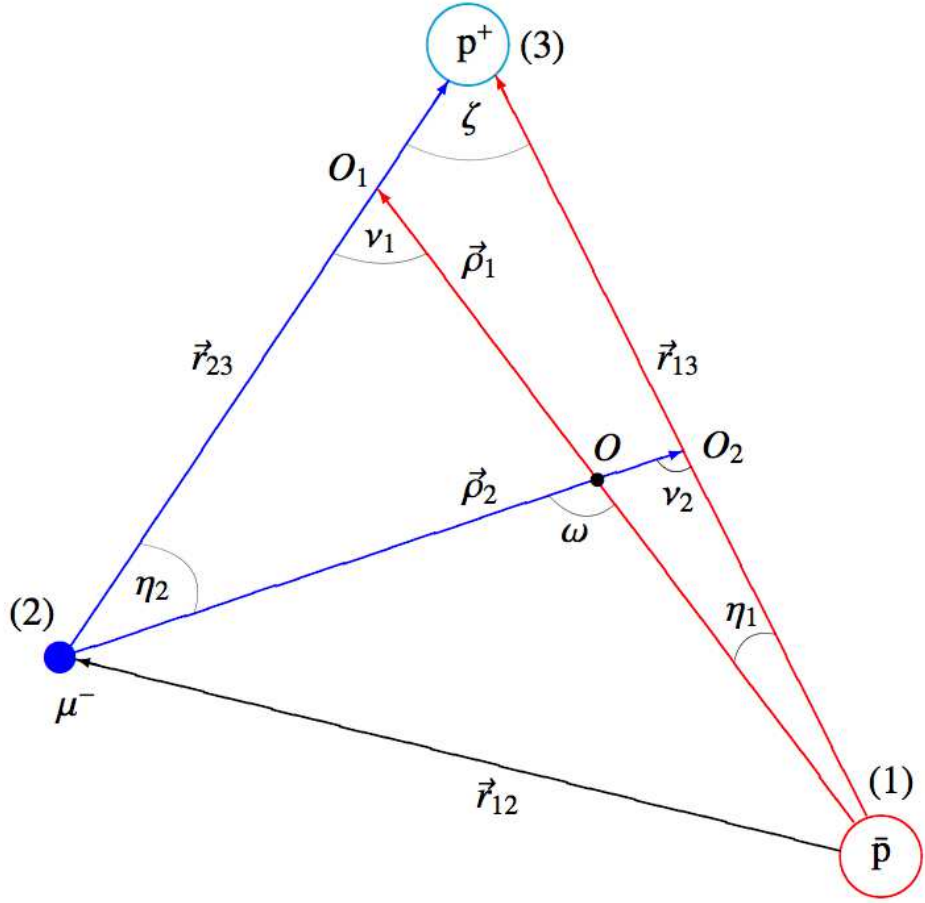}
\caption{The title three-charge-particle system ${\bar{\rm p}},\mu^-$ and p$^+$ (proton)
and system's configurational triangle $\triangle$123 are presented together with the few-body Jacobi coordinates (vectors):
\{$\vec \rho_1, \vec r_{23}$\} and \{$\vec \rho_2, \vec r_{13}$\}. Additionally, $\vec r_{12}$ is the vector between
two negative particles in the system. The necessary geometrical angles between the vectors such as
$\eta_{1(2)}, \nu_{1(2)}, \zeta$ and $\omega$ are also shown in this figure.}
\label{fig2}\end{figure*}

\clearpage
\begin{figure*}
\centering
\includegraphics[width=0.75\textwidth]{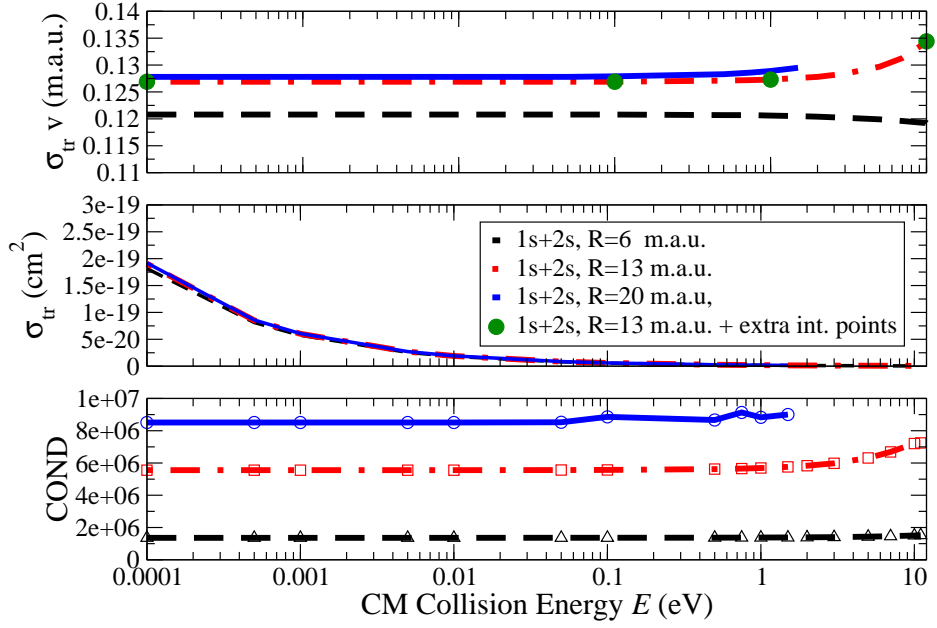}
\caption{Upper plot: numerical convergence results for the
low-energy proton transfer reaction integral cross section $\sigma_{tr}$
multiplied by the collision velocity v $=v_{c.m.}$, i.e. $\sigma_{tr}$v,
in the three-body reaction
$\bar{\rm{p}} + \rm{H}_{\mu}\rightarrow (\bar{\rm{p}}\rm{p})_{\alpha} + \mu^-$.
Here, $ \rm{H}_{\mu}$ is a muonic hydrogen atom and $\alpha=$1s.
Middle plot: same as above but only for the cross section $\sigma_{tr}$.
Lower plot: values of the corresponding COND number (see the text).
In these calculations only the $1s+2s$ MCCA approach is used.
}
\label{fig6}\end{figure*}

\clearpage
\begin{figure*}
\centering
\includegraphics[width=0.75\textwidth]{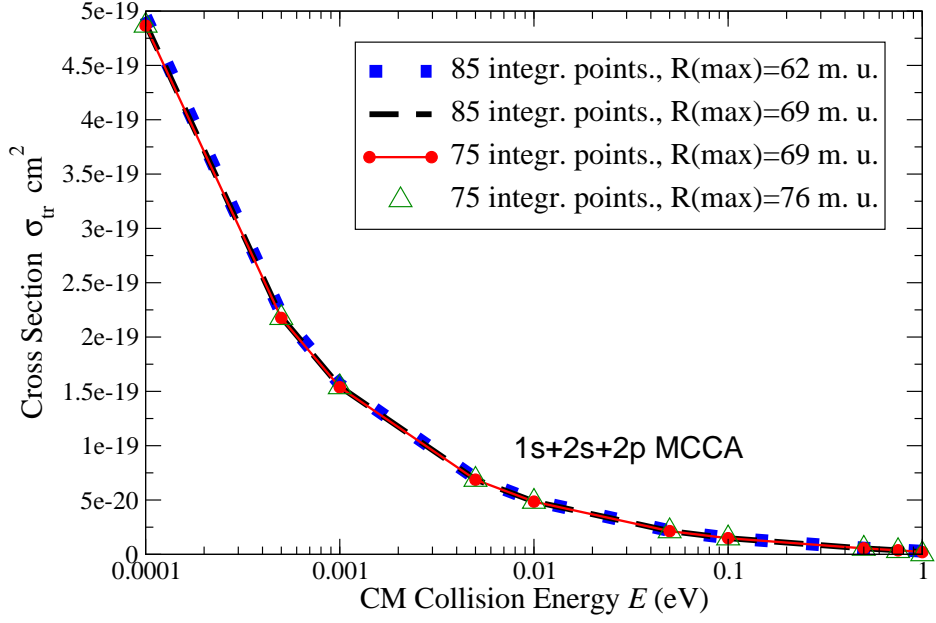}
\caption{Numerical convergence results for the
low-energy proton transfer reaction integral cross section $\sigma_{tr}$ in
$\bar{\rm{p}} + \rm{H}_{\mu}\rightarrow (\bar{\rm{p}}\rm{p})_{\alpha} + \mu^-$, where
$ \rm{H}_{\mu}$ is a muonic hydrogen atom and $\alpha=$1s.
In these calculations the polarization effects are included, i.e. the $1s+2s+2p$ MCCA approach is used.
}\label{fig7}\end{figure*}

\clearpage
\begin{figure*}
\centering
\includegraphics[width=0.75\textwidth]{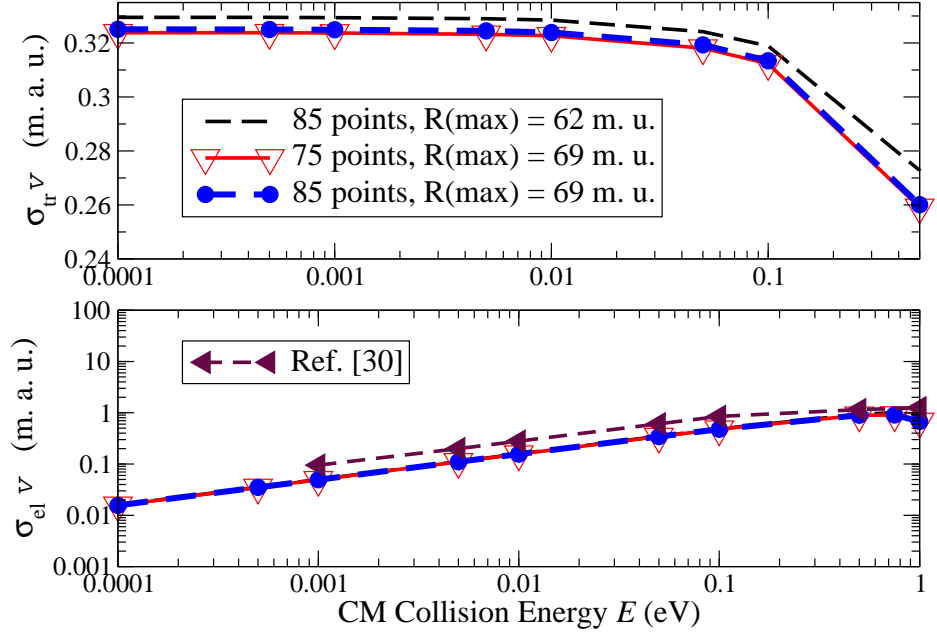}
\caption{Upper plot:
numerical convergence results for the
low-energy proton transfer reaction integral cross section $\sigma_{tr}$
multiplied by the collision velocity v $=v_{c.m.}$.
Lower plot: elastic scattering cross section $\sigma_{el}$ multiplied by the collision velocity v
in $\bar{\rm{p}} + \rm{H}_{\mu}\rightarrow (\bar{\rm{p}}\rm{p})_{\alpha} + \mu^-$, where
$ \rm{H}_{\mu}$ is a muonic hydrogen atom and $\alpha=$1s.
In these calculations only the $1s+2s+2p$ MCCA approach is used.
Corresponding results (triangles left) from paper \cite{igarashi2008} are also shown in this figure.
}
\label{fig8}\end{figure*}

\newpage
\begin{figure*}
\centering
\includegraphics[width=0.75\textwidth]{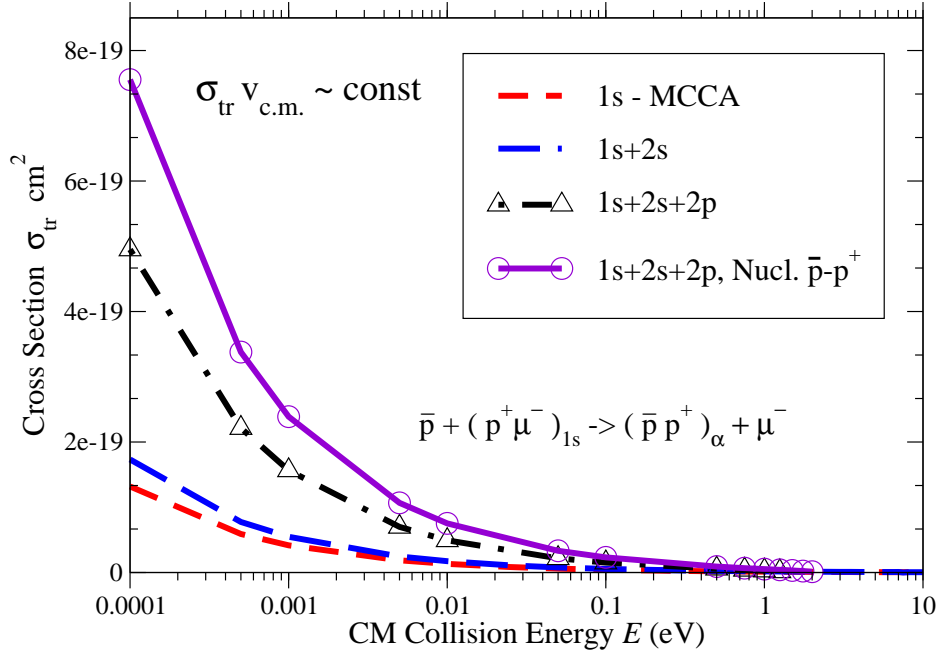}
\caption{This figure shows our final result (after test calculations) for the
low-energy proton transfer reaction integral cross section $\sigma_{tr}$ in the three-charge-particle collision
$\bar{\rm{p}} + \rm{H}_{\mu}\rightarrow (\bar{\rm{p}}\rm{p})_{\alpha} + \mu^-$, where
$ \rm{H}_{\mu}$ is a muonic hydrogen atom: a bound state of a proton and a negative muon. Here only the
reaction's final channel with $\alpha$=1s in considered in the framework of the 1s, 1s+2s
and 1s+2s+2p MCCA approach. The solid line with open circles is the result with an approximate inclusion of the
strong $\bar{\rm{p}}$-$\rm{p}^+$ nuclear interaction.
}\label{fig9}\end{figure*}

\clearpage
\begin{figure*}
\centering
\includegraphics[width=0.75\textwidth]{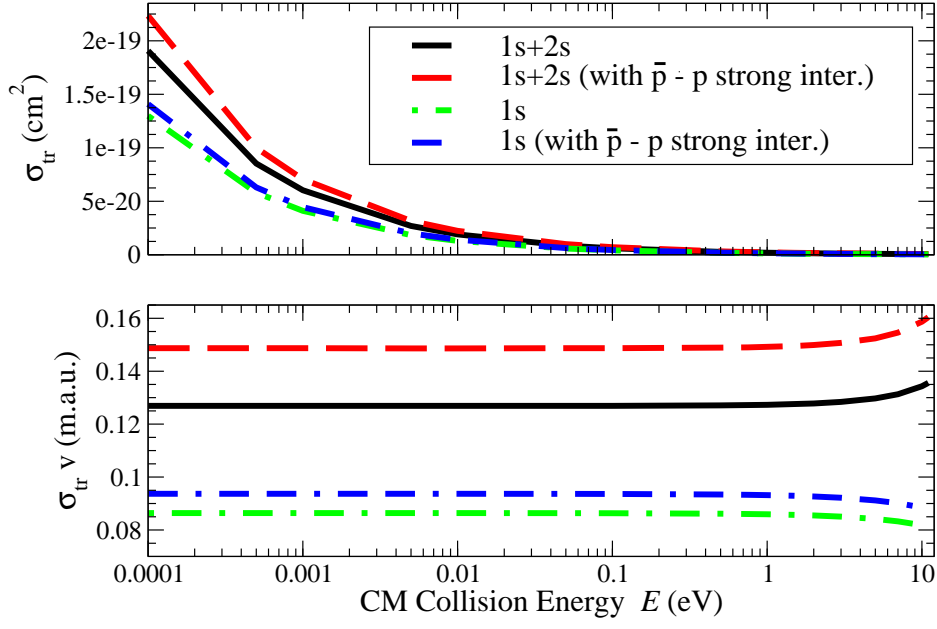}
\caption{Upper plot:
integral cross sections $\sigma_{tr}$ in the reaction (\ref{eq:3}) with and without
inclusion of the $\bar{\rm{p}}$-p strong interaction.
Only the $1s$ and $1s+2s$ approximations are used.
Lower plot: corresponding results as on the top plot, but for the low-energy reaction rate:
$\sigma_{tr}$ multiplied by the collision velocity v $=v_{c.m.}$.
}
\label{fig10}\end{figure*}

\clearpage
\begin{figure*}
\centering
\includegraphics[width=0.75\textwidth]{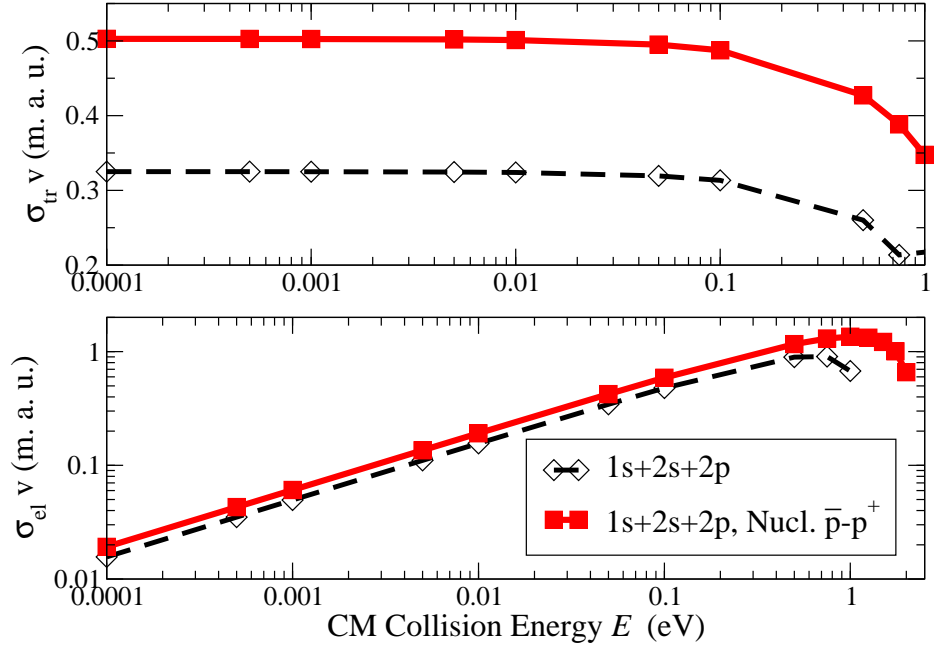}
\caption{Upper plot: the reaction rate, i.e. integral cross sections $\sigma_{tr}$
of the reaction (\ref{eq:3}) multiplied by the collision velocity v
with and without inclusion of the $\bar{\rm{p}}$-p strong interaction for comparison
purposes. Only the $1s+2s+2p$ MCCA method is used in these calculations.
Lower plot: corresponding results as on the top plot, but for the elastic scattering
cross section of the
process (\ref{eq:3}), $\sigma_{el}$, multiplied by the collision velocity v $=v_{c.m.}$.
}
\label{fig11}\end{figure*}

\clearpage
\begin{figure*}
\centering
\includegraphics[width=1.00\textwidth]{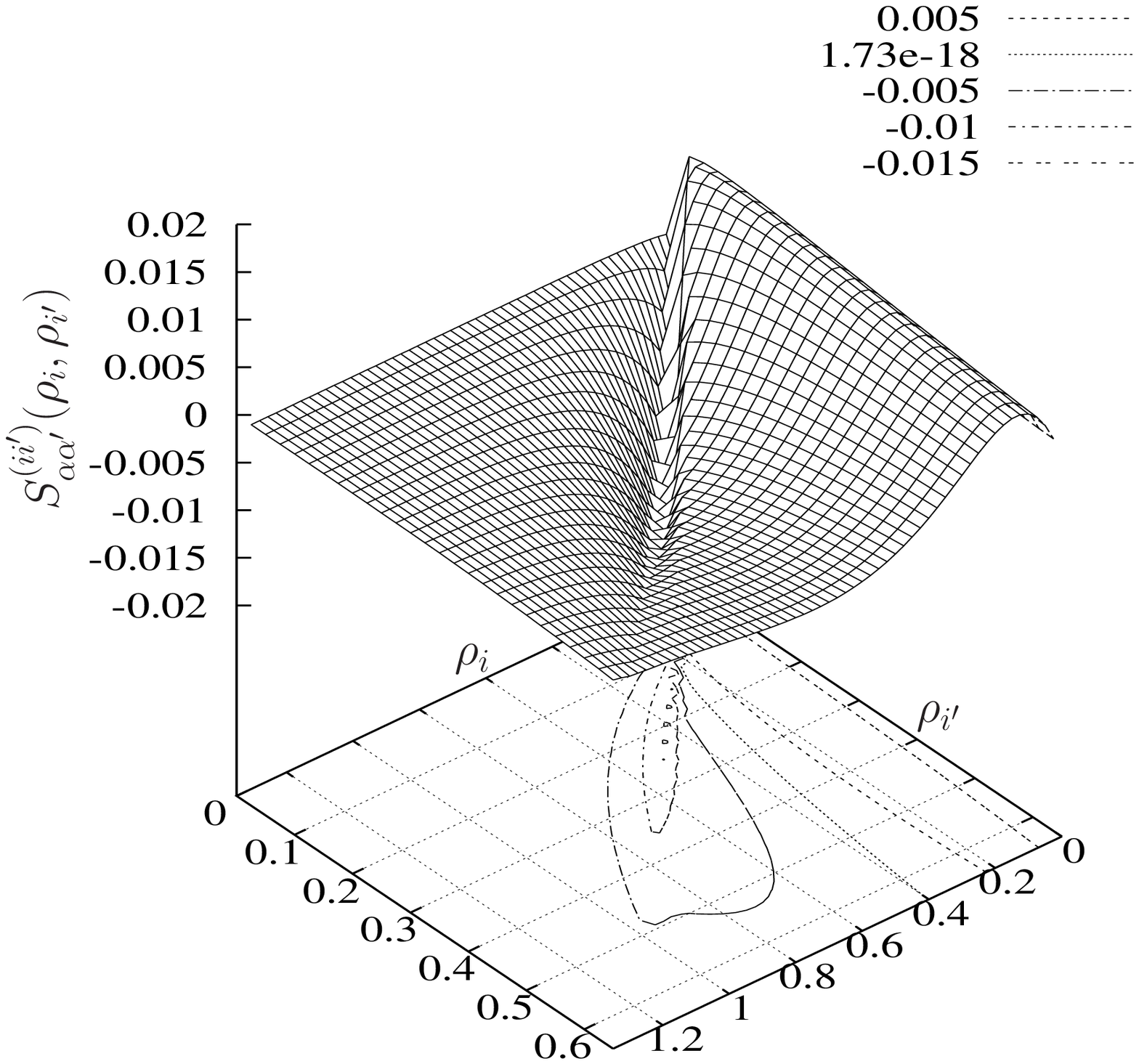}
\caption{The two-dimensional function Eq. (\ref{eq:omega}) 
(three-body angular integral), i.e.
$S_{\alpha \alpha'}^{ii'}(\rho_i, \rho_{i'})$, when $\alpha=\alpha'$=1s.
The values of the coordinates $\{\rho_i, \rho_{i'}\}$ and the surface $S_{\alpha \alpha'}^{ii'}(\rho_i, \rho_{i'})$
are given in atomic muonic units.}
\label{fig3}\end{figure*}

\newpage
\begin{figure*}
\centering
\includegraphics[width=1.00\textwidth]{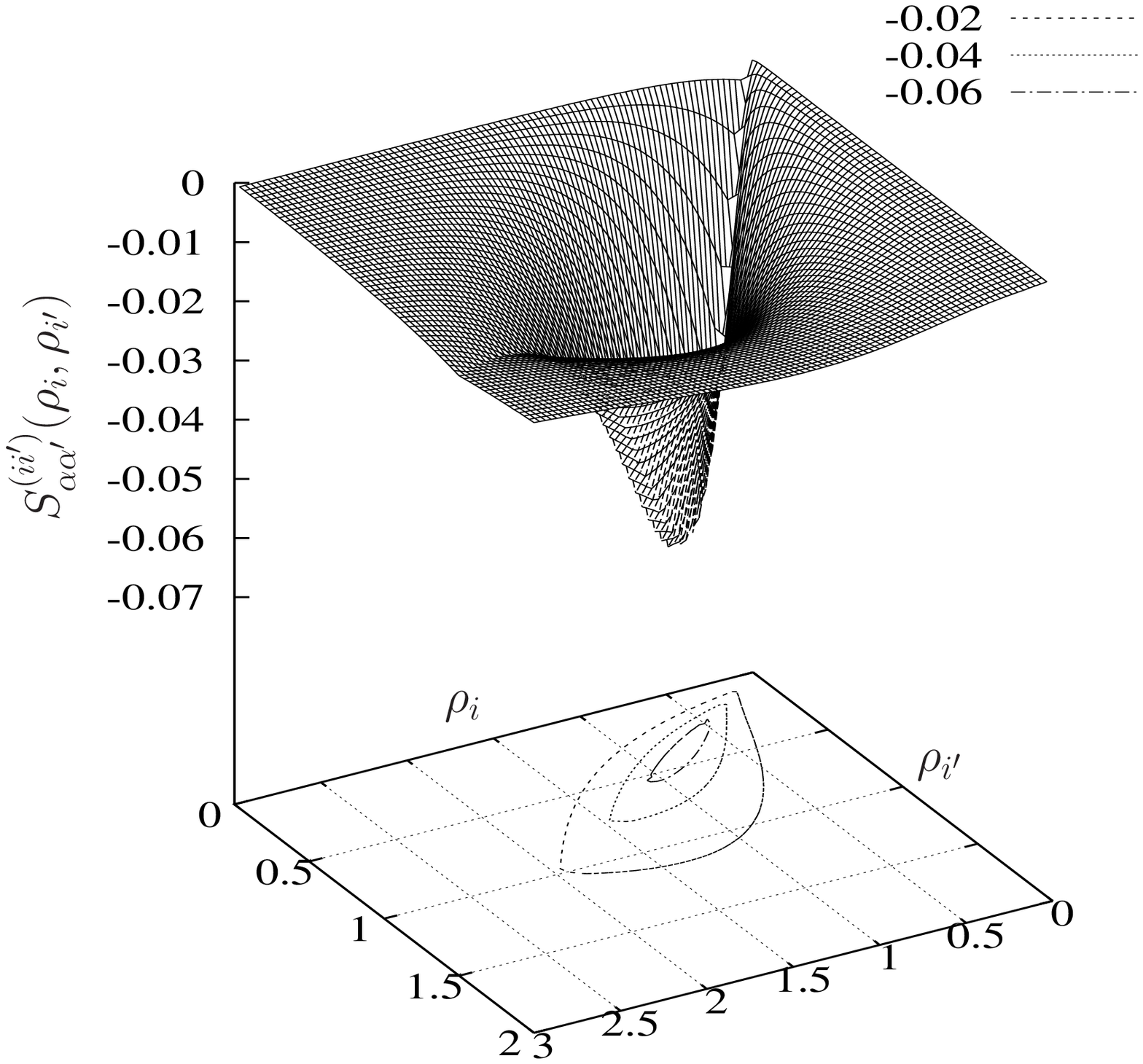}
\caption{The two-dimensional function Eq. (\ref{eq:omega}) 
(three-body angular integral), i.e.
$S_{\alpha \alpha'}^{ii'}(\rho_j, \rho_k)$, when $\alpha$=1s and $\alpha'$=2p.
The values of the coordinates $\{\rho_i, \rho_{i'}\}$ and the surface $S_{\alpha \alpha'}^{ii'}(\rho_i, \rho_{i'})$
are given in atomic muonic units.}
\label{fig4}\end{figure*}

\newpage
\begin{figure*}
\centering
\includegraphics[width=1.00\textwidth]{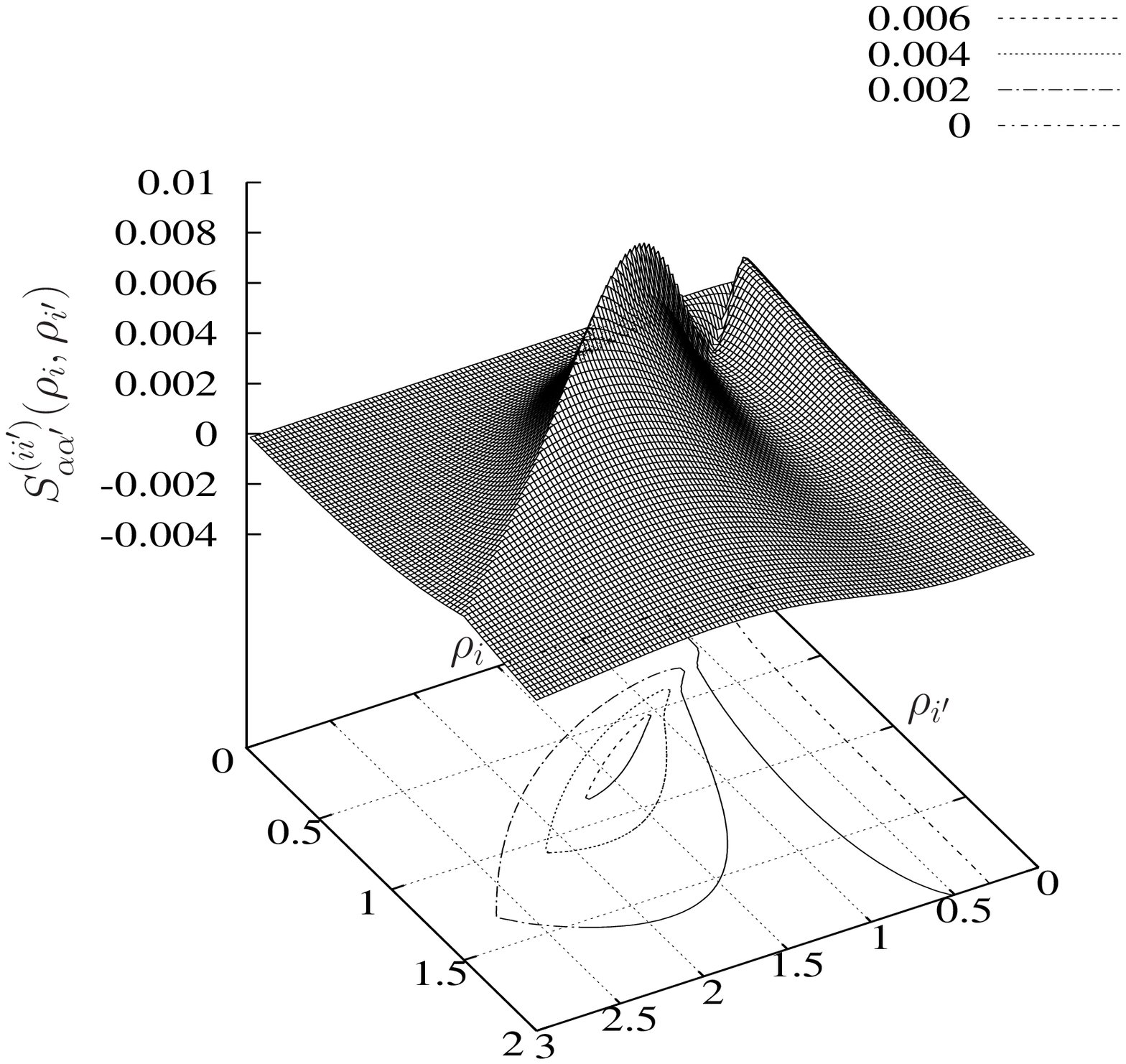}
\caption{The two-dimensional function Eq. (\ref{eq:omega}) 
(three-body angular integral), i.e.
$S_{\alpha \alpha'}^{ii'}(\rho_j, \rho_k)$, when $\alpha=\alpha'$=2s.
The values of the coordinates $\{\rho_i, \rho_{i'}\}$ and the surface $S_{\alpha \alpha'}^{ii'}(\rho_i, \rho_{i'})$
are given in atomic muonic units.}
\label{fig5}\end{figure*}
\end{document}